\documentclass[journal]{IEEEtran}
\usepackage{color} 
\usepackage{algorithm}
\usepackage{graphicx}
\usepackage{cite}
\usepackage{float}
\usepackage{stfloats}
\usepackage{bm}
\usepackage{amsmath,amssymb,amsfonts}
\usepackage{tabularx}
\usepackage{multirow}
\usepackage{threeparttable}
\usepackage{url}

\usepackage[section]{placeins}

\begin{document}

\title{Ultrasonic Backscatter Communication for Implantable Medical Devices
\author{Qianqian Wang,~\IEEEmembership{Graduate Student Member,~IEEE,}
	Quansheng Guan,~\IEEEmembership{Senior Member,~IEEE,}\\
	Julian Cheng,~\IEEEmembership{Senior Member,~IEEE,}
	and~Yuankun~Tang,~\IEEEmembership{Graduate Student Member,~IEEE}}
	\thanks{Q. Wang is with the School of Electronic and Information Engineering, South China University of Technology, Guangzhou 510640,~China, the School of Engineering, The University of British Columbia, Kelowna, BC V1V 1V7, Canada, and also with Key Laboratory of Marine Environmental Survey Technology and Application, Ministry of Natural Resources, Guangzhou 510300, China.}
	\thanks{Q. Guan is with the School of Electronic and Information Engineering, South China University of Technology, Guangzhou 510640,~China, and also with Key Laboratory of Marine Environmental Survey Technology and Application, Ministry of Natural Resources, Guangzhou 510300, China (e-mail: eeqshguan@scut.edu.cn).}
	\thanks{J. Cheng is with the School of Engineering, The University of British Columbia, Kelowna, BC V1V 1V7, Canada.}
	\thanks{Y. Tang is with the School of Electronic and Information Engineering, South China University of Technology, Guangzhou 510640,~China, and also with the Southampton Wireless Group, School of Electronics and Computer Science, University of Southampton, Southampton SO17 1BJ, U.K.}
	}
\maketitle

\begin{abstract}
	 This paper proposes an ultrasonic backscatter communication (UsBC) system for passive implantable medical devices (IMDs) that can operate without batteries, enabling versatile revolutionary applications for future healthcare. The proposed UsBC system consists of a reader and a tag. The reader sends interrogation pulses to the tag. The tag backscatters the pulses based on the piezoelectric effect of a piezo transducer. We present several basic modulation schemes for UsBC by impedance matching of the piezo transducer. To mitigate the interference of other scatters in the human body, the tag transmits information bits by codeword mapping, and the reader performs codeword matching before energy detection in the reader. We further derive the theoretical bit-error rate (BER) expression. Monte Carlo simulations verify the theoretical analysis and show that passive UsBC can achieve low BER and low complexity, which is desirable for size- and energy-constrained~IMDs. 
\end{abstract}


\section{Introduction} \label{sect: introduction}
Intra-body communication (IBC) can enable many applications for future healthcare, such as real-time, continuous, and accurate monitoring for patients with chronic diseases, as well as medical aid for healthy people~\cite{In-body Backcom,D.K. neural,MTPSK}. The major challenge is that IBC has strict constraints on transmission power, device size, and energy consumption since organs are vulnerable and battery replacement in the human body is costly. For instance, although the existing capsule endoscope can make an accurate examination of the gastrointestinal tract, most patients find it challenging to swallow due to its size. Besides, the existing endoscope can only examine the stomach, not the intestines due to the limited battery life~\cite{In-body Backcom}.

Backscatter communication (BC) is an attractive technology, having high energy efficiency \cite{RFID,In-body Backcom,ReofIm}.  A BC system generally has a reader and multiple tags, where the reader transmits interrogation signals, and the tag reflects the signals from the reader to convey the information bits and absorbs energy from the signals to control the logical circuit. Thus, the tag can work as a zero-power device with both communication and energy harvesting functions. Based on the above properties, BC is promising for implantable medical devices (IMDs) to achieve passive IBCs. Specifically, by using BC, the size of the endoscope can be significantly reduced, and the battery can operate for a long period \cite{In-body Backcom}. BC can also be used for brain-computer interface, neural stimulation and recording to treat sleep apnea, diabetes, and rheumatoid arthritis, etc.

The existing research on BC for IMDs commonly adopts radio frequency (RF) waves~\cite{In-body Backcom}. Since the RF wave can be seriously attenuated in the human body and signals in BC systems suffer from round-trip fading, relatively high transmission power is required for RF-based BCs \cite{RFID}. Furthermore, the federal communications commission has a strict intensity limit (10 $\rm mW/cm^2$) on RF waves in the human body \cite{Experimental}. Long-term RF radiation also has potential risks to the human body. Therefore, RF-based BCs are unsuitable for IBCs.

To achieve intra-body BC with low transmission power in the reader, we first propose an ultrasonic BC (UsBC) system, based on the fact that the ultrasonic wave is safe for the human body at most frequencies and power levels and its attenuation is much lower than RF waves. Most importantly, ultrasonic signals have been used for diagnostic and therapeutic purpose for decades. We note that acoustic BC has been realized in underwater acoustic backscatter networks (UABNs) \cite{Underwater}. Thus, UsBC is also feasible in the water-like human body environment. However, the reader in UABN operates at high transmission power (few hundreds of watts) that is impractical to apply for UsBC because UsBC requires much lower power (several $\rm mW$) for the human body safety~\cite{Experimental}. The severe multi-path effect of IBC channels also deserves further research on UsBC for IBCs. To the best of the authors' knowledge, this paper is the first work to study UsBC to enable passive IBCs. 

The proposed UsBC system consists of a reader and a tag and conveys information bits by ultrasonic pulses. The ultrasonic pulse is adopted for its low energy consumption and high resolution to resist the multipath effect in the human body \cite{DSUsWB,UsIM}. The reader transmits interrogation pulses, and the passive tag harvests energy from the pulses and backscatters the pulses to transmit the tag's information bits. Since the intra-body channel is complicated with dense multipath signals, it is challenging to implement coherent receivers that require accurate synchronization between the reader and the tag. Although the non-coherent receiver has low complexity, interference signals that are backscattered from other scatters (i.e., organs and tissues) result in high bit-error rate (BER). The contributions of this paper are summarized as~follows:
\begin{itemize}
	\item We propose a UsBC system for passive IMDs, which is totally different from the traditional active implants that require many power-hungry components, such as oscillators, power amplifiers, and ADCs. We also present several basic modulation schemes for UsBC by impedance matching of the piezo transducer. 
	\item We propose to use block-coded modulation for the tag and perform codeword matching in the reader to pre-mitigate the interference before energy detection (ED), thereby realizing a low-complexity and low BER non-coherent receiver for UsBC.
	\item  We derive the theoretical BER expression over the intra-body fading channels. Extensive Monte Carlo simulation results verify the theoretical analysis and reveal that codeword matching for ED can achieve IBCs with low~BER.
\end{itemize}

The remainder of this paper is organized as follows. Section~II proposes a UsBC system for IBCs. Section~III analyzes system performance. The simulation results are discussed in Section IV, and conclusions are drawn in Section V.

\begin{figure}[!t]
	\centering
	\includegraphics[width=0.7\columnwidth]{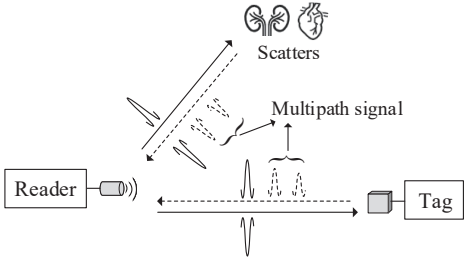}\vspace{-0.3cm}
	\caption{System model of UsBC.}
	\label{System}\vspace{-0.3cm}
\end{figure}

\section{Ultrasonic Backscatter Communication}
The proposed UsBC system consists of a reader, intra-body channels, and a tag, as shown in Fig. \ref{System}. The reader transmits interrogation pulses by an ultrasonic transceiver and decodes the backscattered signals from the tag. The interrogation pulses propagate via intra-body channels, which cause multipath signals. The tag harvests energy from the received pulses to control the impedance matching of the piezo transducer, and performs BC on the interrogation pulses to transmit the tag's bits. The principles are detailed as follows.

\subsection{Backscatter Modulation of Tag} \label{sect: Tag}
We first illustrate the principle of UsBC and then propose the equivalent circuit of the tag to implement several basic modulations, including pulse amplitude modulation (PAM), pulse position modulation (PPM), and binary phase shift keying (BPSK).

For two media with the impedance of $Z_1$ and $Z_2$ in the human body, the ultrasonic signal is backscattered at the boundary. Suppose the signal is transmitted from $Z_1$ and the initial amplitude of the ultrasonic signal is $p_0$, the amplitude of the backscattered signal is expressed as $p_1=p_0(1+RF)$ \cite{Diag.}, where $RF$ is the reflection coefficient, which is given~by
\begin{equation}
	RF = \frac{Z_2-Z_1}{Z_2+Z_1}.
	\label{RF}
\end{equation}
Observing \eqref{RF}, we can derive that if $Z_2\!=\!Z_1$, no reflection occurs and the energy will be~absorbed; if $Z_2\neq Z_1$, there will be a reflection on the ultrasonic wave. Specifically, if $Z_2\!=\!0$, the incident signal will be reflected with a 180-degree inversion; if $Z_2\!=\!\infty$, the incident signal will be reflected.

\begin{figure}[!t]
	\centering
	\includegraphics[width=0.95\columnwidth]{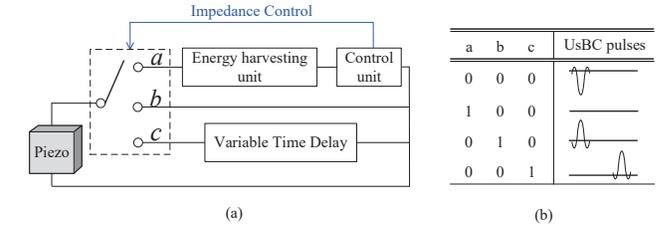}
	\caption{{(a) Equivalent circuit of the tag of UsBC}, and (b) corresponding waveform for turning on each switch.}
	\label{Ciruit}\vspace{-0.3cm}
\end{figure}

Inspired by the aforementioned reflection properties of ultrasonic waves,  by impedance matching of the piezo transducer, we can design different modulation schemes for tags to transmit information bits passively. The equivalent circuit for the tag is shown in Fig.~\ref{Ciruit}(a). Physically, the tag consists of a piezo transducer, an energy harvesting unit, and a control unit. The tag can backscatter signals due to the piezoelectric properties of the piezo transducer. The energy harvesting unit is used to store the energy of the reader's signal and thus the tag does not require a battery. The control unit consumes the harvested energy to control the switches that implement different impedance matching of the piezo transducer. The function of each switch is described as follows:
\begin{itemize}
	\item Off: the pulse is reflected with a 180-degree inversion;
	\item Turn to ``a'': the impedance matches with the piezo transducer, and the energy harvesting unit saves the energy for the control unit of the tag;
	\item Turn to ``b'': the short circuit reflects the pulse;
	\item Turn to ``c'': the variable time delay causes a time delay of the transmitted~pulse.
\end{itemize}
For clarity, the corresponding waveform for turning on each switch are depicted in Fig. \ref{Ciruit}(b), where ``0'' and ``1'' indicate the switch is on and off, respectively. It is easy to find that digital modulations such as on-off keying, PPM, and BPSK can be readily realized. Furthermore, by changing the impedance of the switch ``$a$'', the piezo transducer can reflect pulses with different amplitudes, and thus PAM can be~realized.

\subsection{Channel Model}
We consider the multipath signal in the UsBC system, as shown in Fig. \ref{System}. The human tissue is regarded as an inhomogeneous medium having different densities and sound velocities, and there are numerous organs and particles in the human body~\cite{DSUsWB}. Tissues and organs will attenuate signals and cause reflection and refraction of the transmitted pulses.

\subsubsection{{Channel attenuation}}
{The attenuation of ultrasonic waves in tissues relates to the amplitude attenuation coefficient $\alpha$ (in $\rm [np \cdot cm^{-1}]$) and the transmission distance $d$, and can be calculated as \cite{Experimental} 
\begin{equation}
	\begin{aligned}
		p(d) = p_0 e^{-\alpha d},
	\end{aligned}
\end{equation}
where $p_0$ is the initial amplitude. The parameter $\alpha$ is a function of the central frequency of a channel and $\alpha=af^b$, where $a$ and $b$ are the attenuation parameters that characterize the~tissue.}

{The center frequency of the clinical ultrasound transducer is 1-20 MHz, and thus an appropriate transmission frequency can be selected to alleviate the attenuation effect. Within this frequency range, ultrasonic IBC technology can achieve data rates up to Mbps \cite{Mbps}. Additionally, since the transmission distance of IBC is generally several centimeters, this distance will not dramatically attenuate the ultrasonic signal.}

\subsubsection{Channel fading} Due to the multipath effect of intra-body channels, the channel impulse response is given by~\cite{DSUsWB}
\begin{equation}
	h(t)=\sum_{l=1}^{L}h_l(t)\delta(t-\tau_l),
\end{equation}
where $L$ is the number of multipath signals, $h_l$ and $\tau_l$ denote the path's amplitude and time delay for the $l$th multipath signal, respectively. We assume that the multipath channel is quasi-static, i.e., the amplitude and delay of each path are constant during at least one symbol duration. Resorting to the statistical intra-body channel model that is measured by an ultrasonic software-defined testbed, the fading coefficient $h_l$ follows a generalized Nakagami distribution \cite{Experimental}
\begin{equation}
	\xi(h_l)=\frac{2 s z^{z} h_l^{2 s z-1}}{\Gamma(z) \Omega^{z}} e^{-\frac{z}{\Omega} h_l^{2 s}},
	\label{eqNaka}
\end{equation}
where $z$, $\Omega$ and $s$ are the shaping, spreading, and generalization parameters, respectively, and $\Gamma(\cdot)$ denotes the gamma function.

Since the large instantaneous bandwidth of ultrasonic pulses enables fine time resolution,  the original transmission pulses are resolvable from the multipath signals to overcome the multipath effect \cite{DSUsWB}. To further overcome the interference signals that are backscattered from other scatters to the reader, we next propose an ED reader for the UsBC system.

\begin{figure}[!t]
	\centering
	\includegraphics[width=0.85\columnwidth]{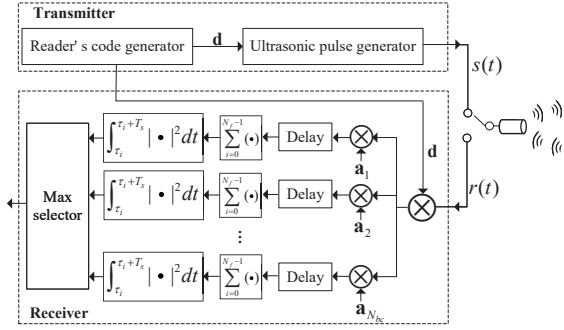}
	\caption{A structure of the reader of UsBC}
	\label{model}
\end{figure}

\subsection{Interference Pre-Mitigation of Reader}
To pre-mitigate the interference and demodulate the tag's signal in the reader, the tag modulates the information bits by codeword mapping and the reader performs codeword matching and ED. The structure of the reader is shown in Fig. \ref{model}, and the communication process includes three phases that are described as follows:

\subsubsection{Signal transmission} The transmitter in the reader first generates the reader's code $\bf d$, where the bold letter represents a vector and elements in $\bf d$ take value from $\{\pm1\}$. Then, the ultrasonic pulse generator periodically transmits a series of pulses having the same amplitude as the reader's code.

\subsubsection{Signal backscatter} The tag implements block-coded modulation \cite{Block} and the information bits are mapped to a codeword in a codebook. The tag controls the amplitude of the backscattered pulses according to the codeword by impedance matching, as described in Section~\ref{sect: Tag}.

\subsubsection{Signal demodulation} The receiver in the reader matches the received signal with all codewords in the codebook and aggregates the signals within each symbol duration. Finally, the reader adopts ED to demodulate the backscattered signals.

Since the codewords are orthogonal, the mapped codeword can be distinguished from other codewords. More importantly, since the codewords are balanced, i.e., each codeword has the same number of ``$+1$'' and~``$-1$'', the zero mean can help eliminate the interference signals from other scatters.

\section{Performance Analysis}
This section analyzes the signal models and derives the theoretical BER expression of UsBC for IBCs.
\subsection{Signal Models}
Considering an infinite sequence of pulses spaced apart by frames, we express the transmitted signal as 
\begin{equation}
	s\left( t \right) = \sum\limits_{i =  - \infty }^\infty  {\sum\limits_{j = 0}^{{N_f} - 1} {d_j p\left( {t - i{T_s} - j{T_f}} \right)} },
	\label{eqtrans}
\end{equation}
where $d_j$ is the $j$th element of the reader's code with the length of $N_f$, $p(t)$ represents the transmitted monocycle waveform having $E_p$ energy, and $T_s$ is the symbol duration with $T_s=N_fT_f$. If the frame time $T_f$ is properly designed, all backscattered signals are received before the transmission of the successive~pulse.

For brevity, we assume that the reader receives the first pulse in the first frame to demodulate the first information symbol of the tag. The compound received signal backscattered from the tag and scatters is given~by
\begin{equation}
	\begin{aligned}
		r(t) = &\sum\limits_{j = 0}^{{N_f} - 1} {{\alpha ^2}{a_{k,j}}{d_j}p\left( {t - j{T_f}} \right)} \\
		&+ \sum\limits_{j = 0}^{{N_f} - 1} {{\beta ^2}{d_j}p\left( {t - j{T_f}} \right)}  + \sum\limits_{j = 0}^{{N_f} - 1}n(t- j{T_f}),
	\end{aligned}
	\label{eqrt}
\end{equation}
where the parameters are described as follows:
\begin{itemize}
	\item $a_{k,j}$ is the $j$th element of ${\bf a}_k$ that is the $k$th codeword in a codebook, and is mapped from the tag's symbol;
	\item $\alpha^2$ represents the round-trip fading of the tag's signal;
	\item ${\beta ^2} = \sum\limits_{l = 1}^S {\beta _{_l}^2}$ denotes the round-trip fading of interference signals backscattered from different $S$ scatters;
	\item $n(t)$ is the additive Gaussian white noise having the first-order distribution of $\mathcal{N}\left(0,\frac{N_0}{2}\right)$, {where $N_0$ is the noise power spectral density}.
\end{itemize}
Since the codebook has $N_{bc}$ balanced orthogonal codewords, each orthogonal codeword can convey $\lfloor\log_2N_{bc}\rfloor$ bits, where $\lfloor \cdot \rfloor$ is the floor function.

To demodulate the tag's bits, as shown in Fig. \ref{model}, the reader multiplies ${\bf a}_{m}$ and $\bf d$ to $r(t)$, where $m=1,\dots,N_{bc}$. {The signal in each frame within each symbol duration is delayed to achieve time alignment. The delayed signals are aggregated in each symbol duration. The aggregated signal is given~by} 
\begin{equation}
	\begin{aligned}
		&y_m(t) = {\bf a}_{m}{\bf d}\cdot r\left( {t + j{T_f}} \right)\\
		&\!=\!\underbrace {{\alpha ^2}\!\!\sum\limits_{j = 0}^{{N_f} \!-\! 1} \!\!{{a_{k,j}}{a_{m,j}}p\left( {t} \right)} }_{{y_{m1}}(t)} \!+\! \underbrace {{\beta ^2}\!\!\sum\limits_{j = 0}^{{N_f} \!-\! 1} \!\!{{a_{m,j}}p\left( {t} \right)} }_{{y_{m2}}(t)} \!+\! \underbrace {\!\!\sum\limits_{j = 0}^{{N_f} \!-\! 1}\!\! {{a_{m,j}}{d_j}n\left( {t} \right)} }_{{n_{m}}(t)}\!.
	\end{aligned}
	\label{eqrt2}
\end{equation}

\emph{Remark} 1:  {It can be observed from \eqref{eqrt2} that the tag's signal $y_{m1}(t)$ is associated with both the tag's codeword ${\bf a}_{k}$ and the matching codeword ${\bf a}_{m}$; while the interference signal $y_{m2}(t)$ is only associated with the matching codeword ${\bf a}_{m}$. To eliminate the interference, we can design the codeword to make $\sum\nolimits_{j=0}^{N_f-1}a_{m,j}=0$.} Therefore, the codewords are set as balanced codes with the same number of ``$+1$'' and~``$-1$''.

After codeword matching, the reader makes ED at the signal to generate the decision variables, i.e.,
\begin{equation}
	\begin{aligned}
		{J_1}\left( {{y_m}(t)} \right) &= \int_{{\tau _1}}^{{\tau_1} + {T_f}} \!\!\!{{{\left| {{y_{m1}}(t) + {n_m}(t)} \right|}^2}} dt,
		\label{eqJym}
	\end{aligned}
\end{equation}
where $\tau_1$ represents the start time of the first symbol duration.

Finally, the symbol can be demapped from the codeword ${\bf{\hat a}}_k$, which is detected according~to 
\begin{equation}
	\begin{aligned}
		{\bf{\hat a}}_k = \arg \mathop {\max }\limits_{m = 1,\dots,{N_{bc}}} \left\{{J_1}\left(y_m(t) \right)\right\}.
	\end{aligned}
\end{equation}

\subsection{BER}\label{Sect: BER}
{Suppose there are $N_{bc}$ codewords in a codebook, $K$ tag's information bits are mapped to a codeword ${\bf a}_k$ in the codebook, where $k\in\left\{1,\dots,N_{bc}\right\}$. The average BER can be calculated~as
\begin{equation}
	P_{e} = \frac{2^{K-1}}{2^K-1}P_{ed},
	\label{eqBER}
\end{equation}
where $P_{ed}$ is the probability of error detection of the codeword. Without loss of generality, suppose the $i$th symbol is transmitted in the tag by the first codeword ${\bf a}_1$, $P_{ed}$ is given~by
\begin{equation}
	\begin{aligned}
		P_{ed}\!=\!1\!-\!\Pr\left({J_i}\left( y_1(t) \right)>\!\!\mathop {\max}\limits_{\substack{l\in\{2,\dots,{N_{bc}}\}}} \{{J_i}\left( y_l(t) \right)\} \right).
		\label{eqPed0}
\end{aligned}
\end{equation}
Let $P_{ed}=1-P_e$, we have
\begin{equation}
	\begin{aligned}
		P_{e}\!=\!\Pr \!\left( {{J_i}\left( {{y_1}(t)} \right) \!>\! {J_i}\left( {{y_2}(t)} \right),\cdots,{J_i}\left( {{y_1}(t)} \right) \!>\! {J_i}\left( {{y_{{N_{bc}}}}(t)} \right)} \right)\!,
		\label{eqPed}
\end{aligned}
\end{equation}
where ${J_i}\left( y_m(t) \right)$, $m= 1,\dots,N_{bc}$, according to \eqref{eqJym}, is given~by 
\begin{equation}
	\begin{aligned}
		{J_i}&\left( {{y_m}(t)} \right) = \int_{{\tau _i}}^{{\tau _i}+ {T_f}} {{{\left| {{y_{m1}}(t)} \right|}^2}} dt \\
			& + 2\int_{{\tau _i}}^{{\tau _i} + {T_f}} {{y_{m1}}(t){n_m}(t)} dt + \int_{{\tau _i}}^{{\tau _i} + {T_f}} {{{\left| {{n_m}(t)} \right|}^2}} dt.
	\end{aligned}
\end{equation}
Specifically,
\begin{equation}
	\begin{aligned}
		{\int_{{\tau _i}}^{{\tau _i} + {T_f}} \hspace{-0.5cm}{\left| {{y_{m1}}(t)} \right|} ^2}dt&\!=\! {\alpha ^4}\int_{{\tau _i}}^{{\tau _i} + {T_f}} \!\!{{{\left( {\sum\limits_{j = 0}^{{N_f} - 1} {{a_{1,j}}{a_{m,j}}p\left( {t} \right)} } \right)}^2\!}dt}\\
		&=\left\{\begin{array}{*{10}{l}}
			{\alpha ^4}N^2_fE_p,\ \  {\bf a}_m={\bf a}_1,\vspace{-0.3cm} \\
			0, \qquad \qquad {\bf a}_m\neq {\bf a}_1.
		\end{array}\right.
	\end{aligned}
	\label{eqy_1m}
\end{equation}
Note that \eqref{eqy_1m} holds because the codewords are orthogonal. When the codeword matches with the received signal, i.e., ${\bf a}_m={\bf a}_1$, we have $\sum\nolimits_{j = 0}^{{N_f} - 1} {a_{1,j}}{a_{1,j}}=N_f$; otherwise, $\sum\nolimits_{j = 0}^{{N_f} - 1} {a_{m,j}}{a_{1,j}}=0$. Furthermore,
\begin{equation}
	\begin{aligned}
		2\int_{{\tau _i}}^{{\tau _i} + {T_f}} {{y_{m1}}(t){n_m}(t)} dt=\left\{\begin{array}{lr}
			n_{1},\ \ \   {\bf a}_m={\bf a}_1,\vspace{-0.3cm} \\
			0, \quad\ \ {\bf a}_m\neq {\bf a}_1,
		\end{array}\right.
	\end{aligned}
\end{equation}
where $n_{1}$ follows the distribution of $\mathcal{N}\left(0,2{\alpha ^4}N^3_fN_0E_p\right)$.
\begin{equation}
	\begin{aligned}
		{\int_{{\tau _i}}^{{\tau _i} + {T_f}}\hspace{-0.4cm}{\left| {{n_m}(t)} \right|} ^2}dt=\eta_m.
	\end{aligned}
	\label{eqn_m}
\end{equation}
It can be derived that ${n_m}(t)$ follows $\mathcal{N}\left(0,\frac{N_fN_0}{2}\right)$. The square-noise term ${\int_{{\tau _i}}^{{\tau _i} + {T_f}}{\left| {{n_m}(t)} \right|} ^2}dt$ can be decomposed into a sum of approximately $2T_fW_{rx}$ independent Gaussian random variables, where $W_{rx}$ is the noise bandwidth that is equal to the bandwidth of the transmitted signal, and its statistical model can be described as a central chi-squared probability density function~\cite{NoncoherentUWB}. Since ultrasonic pulses are wideband signals, it is easy to achieve $T_fW_{rx}>20$. In this case, a central limit theorem applies to the sum so that $\eta_m$ is approximately Gaussian distributed \cite{NoncoherentUWB}, having a mean of $T_fW_{rx}N_fN_0$ and a variance of $T_fW_{rx}N^2_fN^2_0$.} Then, we can re-write \eqref{eqPed} as
\begin{equation}
	\begin{aligned}
		&{P_{e}} \!\approx\! \Pr\! \left( {{\alpha ^4}\!N_f^2{E_p} \!+\! {n_1} \!+\! {\eta _1} \!\!>\! {\eta _2},\!\cdots\!,{\alpha ^4}\!N_f^2{E_p} \!+\! {n_1} \!+\! {\eta _1} \!\!>\! {\eta _{{N_{bc}}}}} \right)\\
		&=  E_{n_x}\left[\prod\limits_{m = 2}^{{N_{bc}}} {\Pr \left( {{\alpha ^4}N_f^2{E_p} + {n_x} > {\eta _m}|{n_x} = {n_1} + {\eta _1}} \right)}\right]\\
		& = E_{n_{x'}}\left[{{\left( {1 - Q\left( {\frac{{{1} + {n_x} - {T_f}{W_{rx}}/\left( {K\gamma } \right)}}{{\sqrt {{T_f}{W_{rx}}/{{\left( {K\gamma } \right)}^2}} }}} \right)} \right)}^{{N_{bc}} - 1}}\right],
		\label{eqPed2}
	\end{aligned}
\end{equation}
where $E_{n_{x'}}\left[\cdot\right]$ represents the expectation with respect to ${n_{x'}}$; ${{n_{x'}}}$ follows ${\mathcal N}\left( {\frac{{{T_f}{W_{rx}}}}{{K\gamma }},\frac{{2}}{{K\gamma }} + \frac{{{T_f}{W_{rx}}}}{{{{\left( {K\gamma } \right)}^2}}}} \right)$; $\gamma=\frac{\alpha^2E_b}{N_0}$ denotes the signal-to-noise ratio (SNR) in the receiver, and $E_b=\frac{N_fE_p}{K}$ is the energy of one bit; $Q(\cdot)$ is the Gaussian $Q$-function. 

Finally, by substituting \eqref{eqPed0} and \eqref{eqPed2} into \eqref{eqBER}, we can obtain the average BER of the UsBC system.

\emph{Remark} 2: It can be deduced that increasing the number of tag's information bits $K$ or increasing $\gamma$ can improve the BER performance. A large $K$ will lead to long codewords, thereby decreasing the data rate. Therefore, a trade-off exists between BER performance and the data rate of UsBC for practical~applications.

\section{Simulation Results} \label{Sect: Sim}
In this section, Monte Carlo simulations are performed to show the BER of the proposed UsBC system over the generalized Nakagami fading channels. {The maximum transmission power of the reader is set to $16\,\mu$W (i.e., SNR $=12$ dB) over a transducer area of 1 $cm^2$ \cite{Experimental}. The parameters of the fading coefficient in \eqref{eqNaka} are $z=0.59$, $\Omega=0.05$, and $s=1.12$, which are measured in a kidney phantom~\cite{Experimental}. The channel attenuation is normalized and the first pulse within each frame is received by the receiver of the reader.}

\begin{figure}[!t]
	\centering
	\includegraphics[width=0.7\columnwidth]{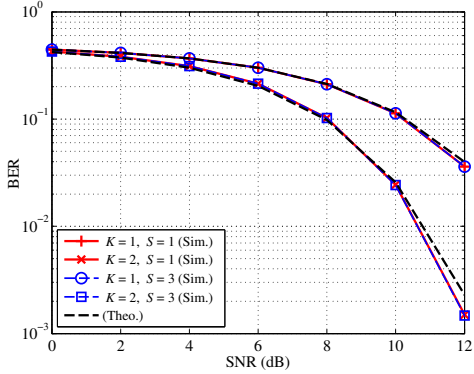}\vspace{-0.3cm}
	\caption{{Theoretical and simulation BERs versus SNR in the cases of the number of tag's bits per symbol $K=1,2$ and the number of scatters~${S}\!=\!1,3$.}}
	\label{BER_SNR}
\end{figure}
\begin{figure}[!t]
	\centering
	\includegraphics[width=0.7\columnwidth]{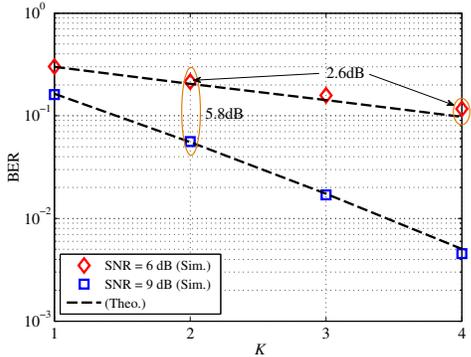}\vspace{-0.3cm}
	\caption{{Theoretical and simulation BERs versus the number of tag's information bits $K$ per symbol in the cases of SNR $=6,9$ dB.}}\vspace{-0.3cm}
	\label{BER_T}
\end{figure}

Fig. \ref{BER_SNR} depicts the theoretical and simulation BERs versus SNR in the cases of the number of tag's bits per symbol $K=1,2$ and the number of scatters ${S}=1,3$. {The theoretical curves become higher than the simulation counterparts at high SNR because the square-noises are approximated as Gaussian variables. The impact of this Gaussian approximation is negligible at low SNR. Since UsBC works at low SNR, the theoretical BER is viable.} In the cases of $K=1$ and ${S}=1,3$, UsBC achieves almost the same BERs, indicating that the interference from other scatters can be effectively eliminated by block-coded modulation in the tag and pre-mitigation in the reader. Thus, \emph{Remark} 1 is proved. Moreover, $K=2$ can achieve a lower BER than $K=1$, indicating that a larger $K$ can improve BER performance, which is consistent with~\emph{Remark~2}.

{Fig. \ref{BER_T} plots the theoretical and simulation BERs versus the number of tag's information bits $K$ per symbol in the cases of SNR $= 6,9$ dB. Take the BER in the case of $K=2$, SNR $=6$~dB as a baseline, we can find that the BER gains are about $2.6$ dB and $5.8$~dB for $K=4$, SNR $=6$ dB and $K=2$, SNR $=9$ dB, respectively.} Thus, increasing the same times of $K$ achieves higher BER compared with increasing the same times of SNR, this is because a larger $K$ results in a longer symbol duration and inevitably leads to greater square-noise of ED.

\section{Conclusions} \label{sect:Conclusions}
This paper proposed a UsBC system to realize passive IBCs. The passive tag harvests and backscatters the ultrasonic pulses from the reader. We proposed to implement codeword mapping in the tag to transmit information bits and use codeword matching in the reader to pre-mitigate interference from other scatters. The reader employs ED to decode the tag's bits. Both theoretical and simulation results show that the UsBC system can achieve passive IBCs with low BER, meanwhile, retain the low complexity of~ED.

Our future work will focus on coding schemes and multi-tag and multi-reader access to achieve passive intra-body~networking. {Encouraged by UABN experiments and the availability of off-the-shelf medical transducers, we will build a prototype to evaluate the performance of this passive IBC technology.}

\end{document}